\documentclass[12pt]{article}
\usepackage{fullpage,epsfig,fancyheadings}
\usepackage[psamsfonts]{euscript}
\usepackage{epic}
\usepackage{amstex}

\begin{document}
\vspace*{-.6in}
\thispagestyle{empty}
\begin{flushright}
CALT-68-2140\\
hep-th/9711029
\end{flushright}
\baselineskip = 20pt

\vspace{.5in}
{\Large
\begin{center}
{\bf The Status of String Theory} \footnote{Lecture
presented at the 31st International Ahrenshoop Symposium
in Buckow, Germany.}

\end{center}}

\begin{center}
John H. Schwarz\footnote{Work supported in part by the U.S. Dept.
of Energy under Grant No. DE-FG03-92-ER40701.}\\
\emph{California Institute of Technology, Pasadena, CA  91125 USA}
\end{center}
\vspace{1in}

\begin{center}
\textbf{Abstract}
\end{center}
\begin{quotation}
\noindent  There have been many remarkable developments in our understanding of
superstring theory in the past few years, a period that has been described as
``the second superstring revolution.''  Several of them are discussed here.
The presentation is 
intended primarily for the benefit of nonexperts.
\end{quotation}
\vfil

\newpage

\pagenumbering{arabic}

\section{Introduction}

This manuscript presents a brief overview of some of the advances in
understanding superstring theory that have been achieved in the last
few years.  It is aimed at physicists who are not
experts in string theory, but who are
interested in hearing about recent developments. 
Where possible, the references cite review papers
rather than original sources.

It is now clear that what had
been regarded as five distinct superstring theories in ten dimensions are better viewed
as five special points in the moduli space of consistent vacua of a single
theory.  Morover, another special limit corresponds to a vacuum with Lorentz invariance
in eleven dimensions.  Some of the evidence that supports this picture is
reviewed in Sect. 2.
The ``second superstring revolution'' is characterized by the discovery of various
non-perturbative properties of superstring theory.  An important aspect of this
is the occurrence of $p$-dimensional excitations, called $p$-branes.  Their
properties are under good mathematical control when they preserve some of the
underlying supersymmetry.  The maximally supersymmetric $p$-branes that occur
in 10 or 11 dimensions are surveyed in Sect. 3.  For detailed reviews of the material
in Sect. 2 and Sect. 3,  see \cite{vafa}  --  \cite{schwarz}.

Sect. 4 describes how
suitably constructed brane configurations can be used to derive, and make more
geometrical, some of the non-perturbative properties of supersymmetric gauge
theories that have emerged in recent years.
Sect. 5 presents evidence for the existence of new
non-gravitational quantum theories in six dimensions.  In particular, there are
pairs of theories with (2,0) and (1,1) supersymmetry that are related by T
duality.  Finally, in Sect. 6, the Matrix Theory proposal, which is a candidate for a
non-perturbative description of M theory in a certain class of
backgrounds, is sketched.  This subject has been reviewed recently in \cite{bilal,banks2}.
This is a rapidly developing subject, which appears likely to be a
major focus of research in the next couple of years.

There have been other interesting developments in the past few years, which are
omitted from this survey.  The most remarkable, perhaps, is the application of
D-brane technology to the study of black hole physics.  This has led to a
microscopic explanation of the origin of black hole thermodynamics in wide
classes of examples.  (For reviews see \cite{maldacena} and \cite{youm}.)
Other omitted topics include applications to particle
physics phenomenology and to cosmology. For two other surveys of
recent developments in string theory, see
\cite{mukhi} and \cite{lerche2}.

\section{M Theory}

A schematic representation of the relationship between the five superstring
vacua in 10d and the 11d vacuum, characterized by 11d supergravity at low
energy, is given in Fig. 1.  The idea is that there is some large moduli
space of consistent vacua of a single underlying theory -- denoted by M here.
The six limiting points, represented as circles, are special in the sense that they are
the ones with (super) Poincar\'e invariance in ten or eleven dimensions.  
The letters on the edges refer to the type of transformation relating a pair of
limiting points. The
numbers 16 or 32 refer to the number of unbroken supersymmetries.  In 10d
the minimal spinor is Majorana--Weyl and has 16 real components, so the
conserved supercharges correspond to just one MW spinor in 
three cases (type I, HE, and HO). Type II
superstrings have two MW supercharges, with 
opposite chirality in the IIA case and the same
chirality in the IIB case.   In 11d 
the minimal spinor is  Majorana with 32 real components. 

\begin{figure}[t]
\centerline{\epsfxsize=5truein \epsfbox{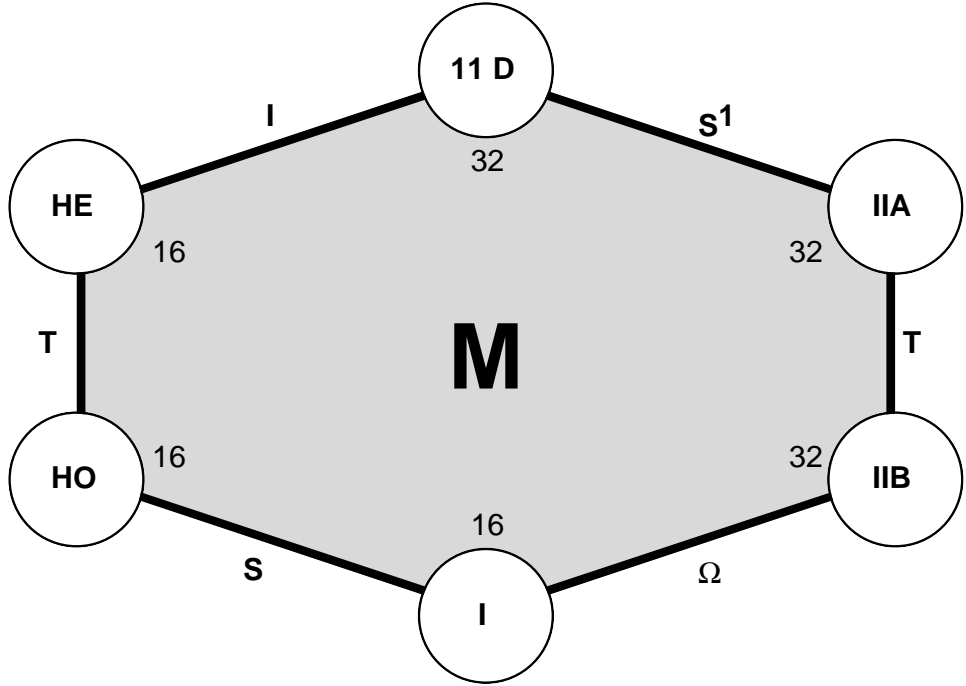} }
\caption{The M  theory moduli space.}
\end{figure}

The 11d vacuum, including 11d supergravity, is characterized by a single scale
-- the 11d Planck scale $m_p$.  It is proportional to $G_N^{-1/9}$, where $G_N$
is the 11d Newton constant.  The connection to type IIA theory is obtained by
taking one of the ten spatial dimensions to be a circle ($S^1$
in the diagram) of radius $R$.
Type IIA string theory in 10d has a dimensionless coupling constant $g_s$, 
which is given by
the vev of $e^\phi$, where $\phi$ is the dilaton field -- a massless scalar
field belonging to the IIA supergravity multiplet.  In addition, the IIA theory
has a mass scale, $m_s$, whose square gives the tension of the 
fundamental IIA string.  The relationship between the parameters  of the
11d and IIA descriptions is given by 
\begin{equation} \label{ems}
m_s^2 = Rm_p^3
\end{equation}
\begin{equation} \label{gs}
g_s = Rm_s.
\end{equation}
Numerical factors (such as $2\pi$) are not important for present purposes and
have been dropped.  The significance of these equations will emerge later.
However, one point can be made immediately.  The conventional perturbative
analysis of the IIA theory is an expansion in powers of $g_s$ with $m_s$ fixed.
 The second relation implies that this is an expansion about $R = 0$, which
accounts for the fact that the 11d interpretation was not evident in studies
of perturbative string theory.  The radius $R$ is a modulus -- the vev
of a massless scalar field with a flat potential.  One gets from the IIA
point to the 11d point by continuing this vev from zero to infinity.  This is
the meaning of the edge of Fig. 1 labeled $S^1$.

The relationship between the $E_8 \times E_8$ heterotic string vacuum (denoted
HE) and 11d is very similar.  The difference is that the compact spatial
dimension is a line interval (denoted I  in the diagram) 
instead of a circle.  The
same relations in eqs. (\ref{ems}) and (\ref{gs}) apply in this case.  
This compactification
leads to an 11d space-time that is a slab with two parallel 10d faces. 
One set of $E_8$ gauge fields is confined to each face, whereas
the gravitational fields reside in the bulk.  There is a nice generalization of
the 10d anomaly cancellation mechanism to this 11d setting \cite{horava}.  It only works for
$E_8$ gauge groups.

The two edges of Fig. 1 labeled T connect vacua related by T duality.
For example, if the IIA theory is compactified on (another) circle of
radius $R_A$ leaving nine noncompact dimensions, this is equivalent to
compactifying the IIB theory on a circle of radius
\begin{equation}
R_B = (m_s^2 R_A)^{-1}.
\end{equation}
Thus, continuing the modulus
$R_A$ from infinity to zero (or $R_B$ from zero to infinity)
gives an interpolation between the IIA and IIB theories.  The T
duality relating the two heterotic
theories (HE and HO) is essentially the same, though there are additional
technical details in this case.

The edge connecting the HO vacuum and the type I vacuum is labeled by S
in the diagram,
since these two vacua are related by S duality.  Specifically, denoting the
two string coupling constants by $g_s^{(HO)}$ and $g_s^{(I)}$, the relation is
\begin{equation}
g_s^{(I)} g_s^{(HO)} = 1.
\end{equation}
In other words, the two dilatons satisfy $\phi^{(I)} + \phi^{(HO)} = 0$, and the
edge connecting the HO and I points in Fig. 1
represents a continuation from weak coupling ($\phi = - \infty$) to strong
coupling ($\phi = + \infty$).  It has been known for a long time that the two
vacua have the same gauge symmetry ($SO(32)$) and the same
supersymmetry, but it was unclear how they could be equivalent
because type I strings and heterotic strings are very different.  The explanation
is that heterotic strings appear as nonperturbative excitations in the
type I description.  The converse is not quite true, because type I strings
disintegrate at strong coupling.

The final link, labeled $\Omega$ in Fig. 1, connects the type IIB and type I
vacua.  $\Omega$ represents an ``orientifold projection,'' which involves modding out
by a particular $Z_2$ discrete symmetry.  Starting from the IIB picture, the
$Z_2$ in question is  an orientation reversal of the IIB string
($\sigma \rightarrow - \sigma$) \cite{sagnotti,horava2}.  
This results in unoriented closed strings
(``untwisted sector'') and unoriented open strings carrying $SO(32)$
gauge symmetry (``twisted sector'').  In the modern viewpoint, the open strings
can be regarded as ending on 32 superimposed D9-branes.  We will say more about
D-branes later.

\section{$p$-branes}

Supersymmetry algebras with central charges admit
``short representations'',  the existence of which is
crucial for testing conjectured
non-perturbative properties of theories that previously were only defined
perturbatively.  Schematically, when a state carries a central
charge $Q$, the supersymmetry
algebra implies that its mass is bounded below $(M\geq |Q|)$.
Moreover, when the state is ``BPS saturated,'' {\it i.e.}, $M = |Q|$, 
the representation theory changes, and a state can
belong to a short representation of the algebra.  This phenomenon is already
familiar for the case of Poincar\'e symmetry in 4d, which 
allows a massless photon to have just two
helicity states (a short representation), 
whereas a massive vector boson must have three helicity
states.

This BPS saturation property arises not only for point particles, characterized
by a mass $M$, but for extended objects with $p$ spatial dimensions, called
$p$-branes.  In this case the central charge is a rank $p$ tensor.  At first
sight, this might seem to be in conflict with the Coleman--Mandula theorem,
which forbids finite tensorial central charges.  However, the $p$-branes carry a
finite charge per unit volume, so that the total charge is infinite for a
BPS $p$-brane that is an infinite hyperplane,
and there is no contradiction.  The BPS saturation condition in this
case implies that the tension (or mass per unit volume) of
the $p$-brane equals the charge density.  Another way of viewing BPS $p$-branes is
as solitons that preserve some of the supersymmetry of the underlying theory.

The theories in question (I will focus on the ones with 32 supercharges) are
approximated at low energy by supergravity theories that contain various
antisymmetric tensor gauge fields.  They are conveniently represented by
differential forms
\begin{equation}
A_n \equiv A_{\mu_{1}\mu_{2} \ldots \mu_{n}} dx^{\mu_{1}} \wedge dx^{\mu_{2}}
\wedge \ldots \wedge dx^{\mu_{n}}.
\end{equation}
In this notation, the corresponding gauge-invariant field strength 
is given by an $(n + 1)$-form $F_{n+1} = dA_n$ plus possible additional terms.  A type
II or 11d supergravity theory with such a gauge field has 
two kinds of BPS $p$-brane solutions, which preserve one-half of the
supersymmetry.  One, which can be called ``electric,'' has $p = n - 1$.  The
other, called ``magnetic,'' has $p = D - n - 3$, where $D$ is the space-time
dimension (ten or eleven for the cases considered here).

A hyperplane with $p$ spatial dimensions in
a space-time with $D-1$ spatial dimensions can be surrounded by a sphere
$S^{D-p-2}$.  If $A$ is a $(p + 1)$-form potential for which a $p$-brane is the
source, the electric charge $Q_E$ of the $p$-brane is given by a 
straightforward generalization of Gauss's law:
\begin{equation}
Q_E \sim \int_{S^{D-p-2}} * F,
\end{equation}
where $S^{D-p-2}$ is a sphere surrounding the $p$-brane and $*F$ is the Hodge
dual of the  $(p + 2)$-form field strength $F$.  Similarly, a dual $(D - p - 4)$-brane
has magnetic charge given by
\begin{equation}
Q_M \sim \int_{S^{p+2}} F.
\end{equation}
The Dirac quantization condition, for electric and magnetic 0-branes in $D =
4$, has a straightforward generalization to a $p$-brane and a dual $(D-p-4)$-brane
in $D$ dimensions
\begin{equation}
{1\over 2\pi} Q_E Q_M \in {\bf Z}.
\end{equation}

An approximate description of the classical dynamics of a ``thin'' $p$-brane is
given by a generalized Nambu--Goto formula
\begin{equation}
S_p = T_p \int \left(\sqrt{-\det G_{\mu\nu}} + \ldots\right) d^{p+1} \sigma,
\end{equation}
where
\begin{equation}
G_{\mu\nu} = g_{MN} (X) \partial_\mu X^M \partial_\nu X^N.
\end{equation}
Here $G_{\mu\nu}(\sigma)$ is a metric on the $(p+1)$-dimension world-volume of
the $p$-brane obtained as a pullback of the D-dimensional space-time metric
$g_{MN}(X)$.  The functions $X^M (\sigma)$ describe the embedding of the
$p$-brane in space-time.  The coefficient $T_p$ is the $p$-brane tension -- its
universal mass per unit volume.  Note that (for $\hbar = c = 1$) $T_p \sim
({\rm mass})^{p+1}$.  This integral is just the volume of the embedded $p$-brane,
generalizing the invariant length of the world-line of a point particle or the
area of the world-sheet of a string.  The dots represent terms involving other
world-volume degrees of freedom required by supersymmetry.

Superstring theories in 10d have three distinct classes of $p$-branes.  These are
distinguished by how the tension $T_p$ depends on the string coupling constant
$g_s$.  A ``fundamental'' $p$-brane has $T_p \sim (m_s)^{p+1}$, with no
dependence on $g_s$.  Such $p$-branes only occur for $p = 1$ -- the fundamental
strings.  Since these are the only objects that survive at $g_s = 0$, they are
the only ones that can be used as the fundamental degrees of freedom in a
perturbative description.  A second class of $p$-branes, called ``solitonic,''
have $T_p \sim (m_s)^{p+1}/g_s^2$.  These only occur for $p = 5$, the
five-branes that are the magnetic duals of the fundamental strings.  This
dependence on the coupling constant is familiar from field theory.  A good example is the
mass of an 't Hooft--Polyakov monopole in gauge theory.  The third class of
$p$-branes, called ``Dirichlet'' (or D$p$-branes), have $T_p \sim
(m_s)^{p+1}/g_s$.  This behavior, intermediate between ``fundamental'' and
``solitonic,'' was not previously known in field theory.  
In 10d type II theories D-branes
occur for all $p\leq 9$ -- even values in the IIA case and odd ones in the IIB
case.  They are all interrelated by T dualities; moreover, the magnetic dual of
a D$p$-brane is a D$p'$-brane with $p' = 6 - p$.  D-branes are very important, and
so we will have more to say about them later.

Eleven-dimensional supergravity contains a three-form potential.  Therefore,
the 11d vacuum admits two basic kinds of $p$-branes -- the M2-brane (also known
as the supermembrane) and the M5-brane.  These are EM duals of one another.
Since the only parameter of the 11d vacuum is the Planck mass $m_p$, their
tensions are necessarily $T_{M2} = (m_p)^3$ and $T_{M5} = (m_p)^6$, up to 
numerical coefficients.

We can use the relation between the 11d theory compactified on a circle of radius
$R$ and the IIA theory in 10d to deduce the tensions of certain IIA $p$-branes.
Starting with the M2-brane we can either allow one of its dimensions to wrap the
circular dimension, leaving a string in the remaining dimensions, or we can
simply embed it in the non-compact dimensions, where it is then still viewed as a
2-brane.  In the latter case, the tension  remains $m_p^3$.  Using eqs. ~(\ref{ems})
and (\ref{gs}), we can recast this as $T = (m_s)^3/g_s$, which we recognize as the
tension of the D2-brane of IIA theory.  On the other hand, the wrapped M2-brane
leaves a string with tension $T = m_p^3 R = m_s^2$.  Thus we see that eq. (\ref{ems})
reflects the fact that a fundamental IIA string is actually a wrapped M2
brane.  Starting with the M5-brane, we can carry out analogous calculations.
If it is not wrapped we obtain a IIA 5-brane with tension $T = m_p^6 =
m_s^6/g_s^2$, which is the correct relation for the solitonic 5-brane
(usually called the NS5-brane).  If it is wrapped on the circle, one is left
with a IIA 4-brane with tension $T = m_p^6 R = m_s^5/g_s$.  This has the
correct tension to be identified as a D4-brane.  In other words, the D4-brane
is actually a wrapped M5-brane.

There are a couple basic facts about D-branes in type II superstring theories
that should be pointed out.  First of all, they can be understood in the weak
coupling limit (which makes them heavy) as surfaces on which fundamental type
II strings can end.  This is where the Dirichlet boundary conditions come in.
This has a number of implications.  One is that the
dynamics of D-branes at weak coupling can be deduced from that of fundamental
strings using perturbative methods.  Another is that since a type II string
carries a conserved charge that couples to a two-form potential, the end of a
string must carry a  point charge, which gives 
rise to electric flux of a Maxwell
field.  This implies that the world-volume theory of a D-brane contains a
$U(1)$ gauge field.  In fact, for strong fields that vary slowly it is actually a
non-linear theory of the Born--Infeld type.  The $U(1)$ gauge field can be
regarded as arising as the lowest excitation of an open string with both ends
attached to the D-brane.

Consider now $k$ parallel D$p$-branes, which are $(p+1)$-dimensional hyperplanes in
${\bf R}^{10}$.  In this case, open strings can end on two different branes.  The
lowest mode of a string connecting the $i$th and $j$th D-brane is a gauge
field that carries  $i$ and $j$ type electric charges at its two ends.
Altogether one has a $U(k)$ gauge theory in $p + 1$ dimensions.
Classically, this can be constructed as the dimensional reduction of $U(k)$
super Yang--Mills theory in 10d.  The separations of the D-branes
are given by the vevs of scalar fields, which break the gauge group to a subgroup.
 For $p \leq 3$, these gauge theories have a straightforward quantum
interpretation, but for $p > 3$ the gauge theories are non-renormalizable.  I will
return to this issue in Sect. 5.

\section{Brane-Configuration Constructions of SUSY Gauge Theories}

In the last section we saw that a collection of $k$ parallel D-branes gives a
supersymmetric $U(k)$ gauge theory.  The unbroken
supersymmetry in this case is maximal (16 conserved supercharges).  In this
section we describe more complicated brane configurations, which
break additional supersymmetries, and give susy gauge theories in 4d with a
richer structure.  This is an active subject, which can be approached in
several different ways.  Here we will settle for two
examples in one particular approach. (For a different approach see \cite{lerche}.)

\begin{figure}[t]
\centerline{\epsfxsize=5truein \epsfbox{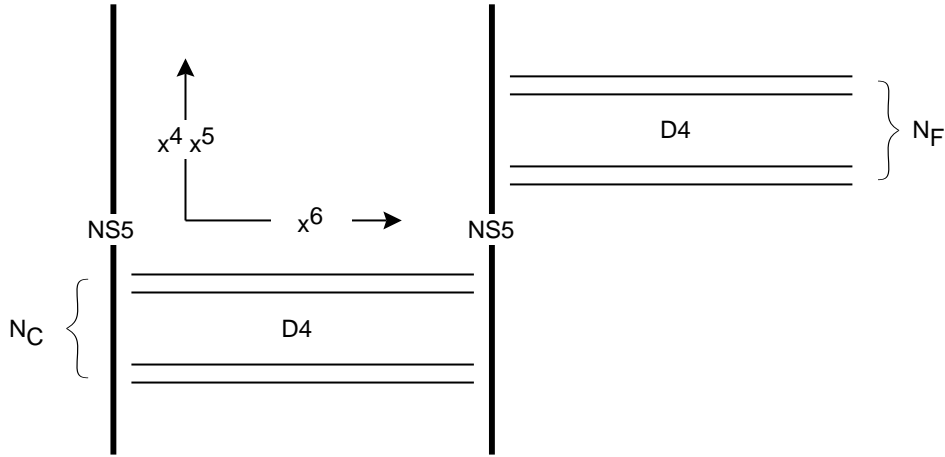} }
\caption{Brane configuration for an $N=2$ 4d gauge theory.}
\end{figure}

The first example \cite{witten1} is a configuration of NS5-branes and D4-branes in type IIA
theory depicted in Fig. 2.  This configuration gives rise
to an $SU(N_C)$ gauge theory in 4d with $N = 2$ supersymmetry (8 conserved
supercharges).  To explain why, one must first describe the geometry.  All of the branes are
embedded in 10d so as to completely fill the dimensions that will be
identified as the 4d space-time with coordinates $x^0, x^1, x^2, x^3$.  In addition, the 
NS5-branes also fill the $x^4$ and $x^5$ dimensions, which are represented by
the vertical direction in the figure, and they have fixed values of $x^6, x^7, x^8,
x^9$.  The D4-branes, on the other hand, have a specified extension in the
$x^6$ direction, depicted horizontally in the figure, and they have fixed values of 
$x^4, x^5, x^7,
x^8, x^9$.  The idea is that the gauge theory lives on the $N_C$
D4-branes, which are suspended between the NS5-branes.  The $x^6$ extension
becomes negligible for 
energies $E \ll 1/L$, where $L$ is the separation between the
NS5-branes.  In this limit the 5d theory on the D4-branes is effectively four
dimensional.  In addition there are $N_F$ semi-infinite D4-branes, which result
in $N_F$ hypermultiplet flavors in the fundamental representation of the gauge
group.  These states arise as the lowest modes of open strings connecting the
two types of D4-branes.  The presence of the NS5-branes is responsible for
breaking the supersymmetry from $N = 4$ to $N = 2$.

This picture is valid at weak coupling, 
because the gauge coupling constant $g_{YM}$ is given by $g_{YM}^2 =
g_s/(Lm_s)$, and the IIA picture is valid for small $g_s$.  Substituting eq.~(\ref{gs}), 
we see that $g_{YM}^2 = R/L$, where $R$ is the radius of a circular
eleventh dimension.  So far, the description of the geometry omits 
consideration of this eleventh
dimension, but by taking it into account we can see what
happens to the gauge theory when $g_{YM}^2$ is not small and quantum effects
become important.  The key step is to recall that a D4-brane is actually an
M5-brane wrapped around the circular eleventh dimension.  Thus, 
reinterpreted as a
brane configuration embedded in 11d, the entire brane configuration corresponds
to a single smooth M5-brane.  The junctions are now smoothed out in a way that
can be made quite explicit.  The correct
configuration is one that is a stable static solution of the M5-brane equation
of motion, which degenerates to the IIA
configuration we have described in the limit $R \rightarrow 0$.
There is a simple method, based on complex analysis, for finding such
solutions.  If space is
described as a complex manifold, with a specific choice of complex structure,
then the brane configuration is a stable static solution if its spatial dimensions
are embedded holomorphically.  In the example at hand, the relevant dimensions
are two dimensions of the M5-brane, which are embedded in the four dimensions denoted
$x^4, x^5, x^6, x^{10}$, where $x^{10}$ is the circular eleventh dimension.  A
complex structure is specified by choosing as holomorphic coordinates $v = x^4
+ i x^5$ and $t = \exp [(x^6 + i x^{10})/R]$, which is single-valued.  Then a
holomorphically embedded submanifold is specified by a holomorphic equation of
the form $F(t,v) = 0$.  The appropriate choice of $F$ is a polynomial in $t$
and $v$ with coefficients that correspond in a simple way to the positions of
the NS5-branes and D4-branes.  (For further details see
Ref. \cite{witten1}.)  This 2d surface is precisely the Seiberg--Witten
Riemann surface (or ``curve'') that characterizes the exact non-perturbative
low-energy effective action of the gauge theory.  When first discovered, this
curve was introduced as
an auxiliary mathematical construct with no evident geometric
significance.  We now see that the Seiberg--Witten solution is given by an
M5-brane with four of its six dimensions giving the space time and the other
two giving the Seiberg--Witten curve!  This simple picture makes the exact
non-perturbative low energy physics of a wide class of $N = 2$ gauge theories
almost trivial to work out.

Let me briefly mention how the brane configuration described above can be
modified to describe certain $N = 1$ susy gauge theories.  One way to achieve
this is to rotate one of the two NS5-branes so that it fills the dimensions
$x^8, x^9$ and has fixed $x^5, x^6$ coordinates.  When this is done the $N_C$
D4-branes running between the NS5-branes are forced to be coincident.  The
rotation breaks the supersymmetry to $N = 1$.  
One of the remarkable discoveries of Seiberg is that an $N =
1$ susy gauge theory with gauge group $SU(N_C)$ and $N_F\geq N_C$ flavors is
equivalent in the infrared to an $SU(N_F - N_C)$ gauge theory with a certain
matter content.  This duality can be realized geometrically in the brane
configuration picture by smoothly deforming the picture so as to move one NS
5-brane to the other side of the other one \cite{elitzur, ooguri}.  
Such a move certainly changes the
exact quantum vacuum described by the configuration.  However, the parameters
involved are irrelevant in the infrared limit, so one achieves a simple understanding
of Seiberg duality.

\section{New Non-gravitational 6d Quantum Theories}

We have seen that it is interesting and worthwhile to consider the world volume
theory of a collection of coincident or nearly coincident branes.  For such a
theory to be regarded in isolation in a consistent way, it is necessary to
define a limit in which the brane degrees of freedom decouple from those of the
surrounding space-time ``bulk.''  Such a limit was implicitly involved in the
discussion of the preceding section.  (This involves some subtleties, which we
did not address.)  In this section we wish to consider the 6d world-volume
theory that lives on a set of (nearly) coincident 5-branes.  If one can
define a limit in which the degrees of freedom of the world-volume theory
decouple from those of the bulk, but still remain self-interacting, then we
will have defined a consistent non-trivial 6d quantum theory \cite{seiberg1}.  
(The only assumption that underlies this is that M theory/superstring theory
is a well-defined quantum theory.)  The 6d quantum theories that
are obtained this way do not contain gravity.  The existence of consistent
quantum theories without gravity in dimensions greater than four came as quite
a surprise to many people.

As a first example consider $k$ parallel M5-branes embedded in flat
11d space-time.  This neglects their effect on the geometry,
which is consistent in the limit that will be considered.  The only
parameters are the 11d Planck mass $m_p$ and the brane separations
$L_{ij}$.  In 11d an M2-brane is allowed to terminate on an M5-brane.
Therefore, a pair of M5 branes can have an M2-brane connect
them.  When the separation $L_{ij}$ becomes small, this M2-brane is well
approximated by a string of tension $T_{ij} = L_{ij} m_p^3$.  The limit that
gives decoupling of the bulk degrees of freedom is $m_p \rightarrow \infty$.
By letting the separations approach zero at the same time, this limit can be
carried out holding the string tensions $T_{ij}$ fixed.  In the limit one
obtains a chiral 6d quantum theory with $(2,0)$ supersymmetry containing $k$
massless tensor supermultiplets and a spectrum of strings with tensions
$T_{ij}$.  There are five massless scalars associated to each brane (parametrizing their
transverse excitations).  They are coordinates for the moduli space of the
resulting theory, which is $({\bf R}^5)^k/S_k$.  The permutation group $S_k$ is due
to quantum statistics for identical branes. String tensions depend on position
in moduli space, and specific ones approach zero at its singularities.

A closely related construction is to consider $k$ parallel NS 5-branes in the
IIA theory.  The difference in this case is that one of the transverse
directions (parametrized by one of the five scalars) is the circular eleventh
dimension.  In carrying out the decoupling limit one can send the radius $R$ to
zero at the same time, holding the fundamental type IIA string tension $T =
m_s^2 = m_p^3 R$ fixed.  The resulting decoupled 6d theory
contains this string in addition to the ones described above.  It becomes
bound to the NS5-branes in the limit, as the amplitude to come free
vanishes in the limit $g_s \rightarrow 0$.  The resulting theory has the moduli
space $({\bf R}^4 \times S^1)^k/S_k$.
This theory contains fundamental strings and has a chiral extended
supersymmetry, features that are analogous to type IIB superstring theory in
10d.  However, it is actually a class of non-gravitational theories (labeled by
$k$) in 6d.  Because of the analogy some authors refer to this
class of theories as $iib$ string theories.  Six-dimensional non-gravitational
analogs of type IIA string theory, denoted $iia$ string theories, are obtained
by means of a similar decoupling limit applied to a set of parallel NS5-branes
in IIB theory.  These $iia$ and $iib$ string theories are related by T
duality.  Explicitly, compactifying one spatial dimension on a circle of radius
$R_a$ or $R_b$, the theories (with given $k$) become equivalent for the
identification $m_s^2 R_a R_b = 1$.  This feature is directly inherited from
the corresponding property of the IIA and IIB theories.  

There are various
generalizations of these theories that will not be described here.  There are
also 6d non-gravitational counterparts of the two 10d heterotic theories.
These have chiral ($1,0$) supersymmetry.  In the notation of Fig. 1, they could
be referred to as $he$ and $ho$ theories.  They, too, are related by T
duality. Although the constructions make us confident about the existence and certain
general properties of these theories, they are not very well understood.  The
10d string theories have been studied for many years, whereas these 6d string
theories are only beginning to be analyzed.  Like their 10d counterparts, the
fact that they have T dualities implies that they are not
conventional quantum field theories.

\section{The Matrix Theory Proposal}

The discovery of string dualities and the connection to 11d
has taught us a great deal about non-perturbative properties of
superstring theories, but it does not constitute a complete non-perturbative
formulation of the theory.  In October 1996, Banks, Fischler, 
Shenker, and Susskind made a
specific conjecture for a complete nonperturbative definition of the theory in
eleven uncompactified dimensions called `Matrix Theory' \cite{banks1,banks2}.  
In this approach, as we will see, other compactification
geometries require additional inputs.  It is far from obvious that the
BFSS proposal is well-defined and
consistent with everything we already know.  
However, it seems to me that there is
enough that is right about it to warrant the intense scrutiny that it has
received and is continuing to receive.  At the time of this writing,
the subject is in a state of turmoil. On the one hand,
there is a new claim that the BFSS prescription (as well as a variant due to
Susskind \cite{susskind}) can be derived from previous knowledge \cite{seiberg2}.  
On the other, some people \cite{dine,douglas,becker} are (cautiously)
claiming to have found specific settings in which it gives wrong answers!  In
the following, we do not comment further upon these claims.  Instead, we describe the
basic ideas of Matrix Theory, as well as some of its successes and limitations.

One of the $p$-branes that has not been discussed yet is the D0-brane of type IIA
theory in 10d.  Being a D-brane, its mass is $M = m_s/g_s$.  Using
eq.~(\ref{gs}), one sees that $M = 1/R$, which means that it can be understood as the
first Kaluza--Klein excitation of the 11d supergravity multiplet on the circular
eleventh dimension.  In fact, this is a good way of understanding (and
remembering) eq.~(\ref{gs}).  Like all the type II D-branes it is a BPS state that
preserves half of the supersymmetry, so one has good mathematical control.
{}From the 11d viewpoint it can be viewed as a wave going around the eleventh
dimension with a single quantum of momentum.  Higher Kaluza--Klein excitations
with $M = N/R$ are also BPS states.  From the IIA viewpoint 
these are bound states of $N$ D0-branes with zero
binding energy.  The existence of a bound state at a threshold is a very subtle
dynamical question, which must be true in this case.  This has in fact been proved
for $N=2$ in ref. \cite{sethi} and for all prime values of $N$ in \cite{porrati}.

By the prescription given in Sect. 4,
the dynamics of $N$ D0-branes is described by the dimensional reduction of $U(N)$
super Yang--Mills theory in 10d to one time dimension only.  When this is done,
the spatial coordinates of the $N$ D0-branes are represented by $N\times N$
Hermitian matrices! This theory has
higher order corrections, in general.  However, one can speculate that these
effects are suppressed by viewing the $N$ D0-branes in the infinite momentum
frame (IMF).  This entails letting $p_{11} = {N/ R}$ approach infinity at
the same time as $R \rightarrow \infty$.  The techniques involved here are
reminiscent of those developed in connection with the parton model of hadrons
in the late 1960's.  The BFSS conjecture is that this IMF frame $N \rightarrow
\infty$ limit of the D0-brane system constitutes an exact non-perturbative
description of the 11d quantum theory.  The $N \rightarrow \infty$ limit is
awkward, to say the least, for testing this conjecture.  A stronger version of
the conjecture, due to Susskind, is applicable to finite $N$.  It asserts that
the IMF D0-brane system, with fixed $N$, provides an exact non-perturbative
description of the 11d theory compactified on a light-like circle with $N$
units of (null) momentum along the circle.

One of the first issues to be addressed was how this conjecture should be
generalized when additional dimensions are compact, specifically if
they form an $n$-torus $T^n$.  The reason this is a non-trivial problem is that
open strings connecting pairs of D0-branes can lie along many topologically
distinct geodesics.  It turns out that all these modes can be taken into account
very elegantly by replacing the 1-dimensional quantum theory of the D0-branes
by an $(n + 1)$-dimensional quantum theory, where the $n$ spatial dimensions
lie on the dual torus $\tilde T^n$.  The extra dimensions precisely account for
all the possible stretched open strings.  This picture had some immediate
successes.  For example, it nicely accounted for all the duality symmetries for
various values of $n$.  However, $(n + 1)$-dimensional super Yang--Mills theory
is non-renormalizable for $n>3$, so this description of the theory is certainly
incomplete in those cases.  The new theories described in
Sect. 5 provide natural candidates when $n = 4$ or $5$, but when $n>5$
there are no theories of this type, and so we seem to be stuck.

In conclusion, Matrix Theory is a very interesting proposal for defining M
theory non-perturbatively.  Whether it is correct, or needs to be modified, is
very much up in the air at the present time.  However, even if it is right, it
is unclear how to define vacua with more than five compact dimensions.
This fact is very intriguing, since this is precisely what is required to
describe the world that we observe.

\end{document}